# A field theory approach to the evolution of canonical helicity and energy

S. You

*William E. Boeing Department of Aeronautics & Astronautics, University of Washington, Seattle, Washington 98195, USA*

A redefinition of the Lagrangian of a multi-particle system in fields reformulates the single-particle, kinetic, and fluid equations governing fluid and plasma dynamics as a single set of generalized Maxwell's equations and Ohm's law for canonical force-fields. The Lagrangian includes new terms representing the coupling between the motion of particle distributions, between distributions and electromagnetic fields, with relativistic contributions. The formulation shows that the concepts of self-organization and canonical helicity transport are applicable across single-particle, kinetic, and fluid regimes, at classical and relativistic scales. The theory gives the basis for comparing canonical helicity change to energy change in general systems. For example, in a fixed, isolated system subject to non-conservative forces, a species' canonical helicity changes less than total energy only if gradients in density or distribution function are shallow.

## I. INTRODUCTION

Self-organization is concerned with the spontaneous emergence of large-scale structures in physical systems. A fundamental conjecture, borrowed from the mathematics of topology [1], is the invariance of a global property during the process of self-organization: for example, a system relaxes to reduce energy but is constrained by a constant value of the helicity of a vector field. Above a critical threshold, the system then forms a large-scale structure of that specific vector field. Hydrodynamic flow helicity models [2,3] have been applied to the forecasting of tornadoes [4], knotting of DNA [5], entanglement of polymers [6], and wing-tip vortices [7]. Magnetic field helicity models are the foundation of theories for the origin of cosmic magnetic fields 8, astrophysical jets [9], solar coronal loops [10], and toroidal magnetic confinement concepts [11, 12]. Arguments ranging from maximal entropy [13] to selective decay [14] attempt to justify why helicity is conserved while energy is minimized [15], leading to magnetic- [15], neutral- [3], or at best, multi-fluid [14, 16, 17] relaxation models. Earlier work has demonstrated an isomorphism between two-fluid plasmas and Maxwell's equations [18] and investigated the helicity of a fluid in a relativistic context [19]. A severe limitation is that all these theories rest on simple and restrictive descriptions of fluids and plasmas: Euler equations for fluids, magnetohydrodynamic or barotropic multi-fluid equations for plasmas. This paper shows that the fundamental transport equation governing helicity evolution is valid across all classical field theory, including relativistic, single particle, kinetic, and fluid regimes. The framework takes into account dissipation, collisionless situations, collective behavior, particle reactions and electromagnetic interactions. This new formulation is derived directly in a Lagrangian-Hamiltonian framework and results in a canonical form of the equation of motion expressed as an Ohm's law and Maxwell's equations for canonical fields. This field theory approach shows that in a simple dissipative system, if the density gradient is weak, helicity changes more slowly than total energy, but if the density gradient is large, helicity changes more rapidly than total energy. This is a first principles explanation for the ruggedness of helicity invariants with respect to energy conservation, and provides a criterion for determining where and when constrained relaxation is applicable.

The paper is organized as follows. Section II presents the gauge-invariant relative canonical helicity transport equation and explains that it is based on the chosen equation of motion of the system. Earlier work has always considered a version of the fluid equations of motion which can be written in the form of an Ohm's law. Section III derives the Maxwell (field) equations from these equations of motion. Section IV shows that the single-particle and kinetic equations can be re-arranged into the same Ohm's law form. Section V uses the Lagrangian formalism to fully generalize the field theory framework. Section VI compares the helicity evolution to the total energy evolution in the simple, isolated, dissipative case, before concluding in Section VII.

| | Particle regime | Kinetic regime | Fluid regime |
|---|---|---|---|
| Canonical momentum $\vec{P}_\sigma$ | Definition: $\vec{P}_\sigma^{par} \equiv \gamma m_\sigma \vec{v}_\sigma + q_\sigma \vec{A}$ <br><br> Newtonian limit: $\vec{P}_\sigma^{par} = m_\sigma \vec{v}_\sigma + q_\sigma \vec{A}$ | Definition: $\vec{P}_\sigma^{kin} \equiv f_\sigma \vec{P}^{par}$ <br><br> Newtonian limit: $\vec{P}_\sigma^{kin} = f_\sigma m_\sigma \vec{v}_\sigma + f_\sigma q_\sigma \vec{A}$ | Definition: $\vec{P}_\sigma^{flu} \equiv \int \vec{P}_\sigma^{kin} d\vec{v}_\sigma$ <br><br> Newtonian limit: $\vec{P}_\sigma^{flu} = \rho_\sigma \vec{u}_\sigma + \rho_{c\sigma} \vec{A}$ |
| Enthalpy $h_\sigma$ | Definition: $h_\sigma^{par} \equiv \gamma m_\sigma c^2 + q_\sigma \phi$ <br><br> Newtonian limit: $h_\sigma^{par} = \frac{m_\sigma v_\sigma^2}{2} + q_\sigma \phi$ | Definition: $h_\sigma^{kin} \equiv f_\sigma h_\sigma^{par}$ <br><br> Newtonian limit: $h_\sigma^{kin} = f_\sigma \frac{m_\sigma v_\sigma^2}{2} + f_\sigma q_\sigma \phi$ | Definition: $h_\sigma^{flu} \equiv \int h_\sigma^{kin} d\vec{v}_\sigma$ <br><br> Newtonian limit: $h_\sigma^{flu} = \rho_{c\sigma} \phi + \frac{1}{2}\rho_\sigma u_\sigma^2 + \mathcal{P}_\sigma$ |
| Canonical vorticity $\vec{\Omega}_\sigma \equiv \nabla \times \vec{P}_\sigma$ | Newtonian limit: $\vec{\Omega}_\sigma^{par} = m_\sigma \vec{\omega}_\sigma^{par} + q_\sigma \vec{B}$ <br><br> where $\vec{\omega}_\sigma^{par} = \nabla \times \vec{v}_\sigma$ | Newtonian limit: $\vec{\Omega}_\sigma^{kin} = f_\sigma \vec{\Omega}_\sigma^{par} + \nabla f_\sigma \times \vec{P}_\sigma^{par}$ | Newtonian limit: $\vec{\Omega}_\sigma^{flu}$ <br> $= \rho_\sigma \vec{\omega}_\sigma^{flu} + \rho_{c\sigma} \vec{B} + \nabla n_\sigma$ <br> $\times (m_\sigma \vec{u}_\sigma + q_\sigma \vec{A})$ <br><br> where $\vec{\omega}_\sigma^{flu} = \nabla \times \vec{u}_\sigma$ |
| Canonical force-field $\vec{\Sigma}_\sigma \equiv -\nabla h_\sigma - \frac{\partial \vec{P}}{\partial t}$ | Newtonian limit: $\vec{\Sigma}_\sigma^{par}$ <br> $= q_\sigma \vec{E} - \nabla\left(\frac{m_\sigma v_\sigma^2}{2}\right)$ <br> $- m_\sigma \frac{\partial \vec{v}_\sigma}{\partial t}$ | Newtonian limit: $\vec{\Sigma}_\sigma^{kin}$ <br> $= f_\sigma \vec{\Sigma}_\sigma^{par} - h_\sigma^{par} \nabla f_\sigma$ <br> $- \vec{P}_\sigma^{par} \frac{\partial f_\sigma}{\partial t}$ | Newtonian limit: $\vec{\Sigma}_\sigma^{flu}$ <br> $= \rho_{c\sigma} \vec{E} - \nabla\left(\frac{\rho_\sigma u_\sigma^2}{2} + \mathcal{P}_\sigma\right)$ <br> $- \frac{\partial (\rho_\sigma \vec{u}_\sigma)}{\partial t}$ |

**Table 1:** Definitions of the canonical momentum and enthalpy of the species $\sigma$ with the resulting canonical fields. The species particles have mass $m_\sigma$, charge $q_\sigma$, a velocity distribution function $f_\sigma(\vec{v}_\sigma)$, which could be integrated as usual to the fluid regime with mass density $\rho_\sigma$, the charge density $\rho_{c\sigma}$, flowing at the bulk velocity $\vec{u}_\sigma$ with pressure $\mathcal{P}_\sigma$ (see text). Dissipative terms are defined in Section IV and more rigorously derived in Section V. The electromagnetic field is defined by the electrostatic potential $\phi$ and magnetic potential $\vec{A}$. The symbols $\gamma(\vec{v}_\sigma)$ and $c$ represent the Lorentz factor and the speed of light, respectively.

## II. CANONICAL HELICITY EVOLUTION

Each species $\sigma$ present in a system has a specific canonical helicity, defined [20] as the volume integral $K_\sigma \equiv \int \vec{P}_{\sigma-} \cdot \vec{\Omega}_{\sigma+} \, dV$ of the canonical vorticity $\vec{\Omega}_\sigma \equiv \nabla \times \vec{P}_\sigma$, which is the circulation of the canonical momentum $\vec{P}_\sigma$ as defined in Table 1. The positive and negative subscripts refer to reference fields (e.g. for vectors $\vec{X}_\pm \equiv \vec{X} \pm \vec{X}_{ref}$, for scalars $x_\pm \equiv x \pm x_{ref}$) so $K_\sigma$ remains gauge invariant even if the canonical fields intercept the boundaries of the species sub-volume. The evolution of the species' specific canonical helicity is given by the transport equation [20]

$$\frac{dK_\sigma}{dt} = 2\int \left(\vec{R}_\sigma \cdot \vec{\Omega}_\sigma\right)_- dV + \int h_{\sigma-}\, \vec{\Omega}_\sigma \cdot d\vec{S} + \int \vec{P}_{\sigma-} \times \frac{\partial \vec{P}_{\sigma+}}{\partial t} \cdot d\vec{S} + \int \vec{P}_{\sigma-} \cdot \vec{\Omega}_{\sigma+}\, \vec{u}_\sigma \cdot d\vec{S}. \quad (1)$$

Eq. 1 includes decay terms due to dissipative forces $\vec{R}_\sigma$, transfer terms due to enthalpy $h_\sigma$ on the boundaries, inductive terms due to time varying $\partial \vec{P}_\sigma/\partial t$, and changes due to boundary motions $\vec{u}_\sigma$. The exact form of $\vec{R}_\sigma$ will be determined in section IV. Eq. 1 is a generalization of earlier, more restrictive, helicity evolution equations [3, 14, 15]: e.g. magnetic helicity evolution is retrieved in the limit of zero inertia ($m_\sigma \to 0$) and the fluid kinetic helicity evolution is retrieved in the limit of zero charge ($q_\sigma \to 0$). Eq. 1 governs the interaction between flow fields of multiple species, or between magnetic fields and flow fields. The method of interaction is helicity transfer between species, or between magnetic fields and a given species [20, 21]. For example, the criterion that determines whether helicity is transferred into a magnetic component or into a species flow component is the skin depth of the species normalized to a scale length of the system [20]. This criterion provides a first principles explanation for the bifurcation threshold measured in a merging magnetized plasma experiment [22]. The twisting and interlinking of canonical flux tubes gives a geometric interpretation of the evolution of canonical helicity; of the interaction between flows in multi-species systems; and of the coupling between flows and magnetic fields in magnetized plasmas [21]. Critically, the helicity transport equation Eq. 1 rests entirely on the chosen equation of motion governing the evolution of the system. The historical choices have been the single magnetohydrodynamic fluid equation of motion [15], the neutral fluid equation of motion [3], or more recently the multi-fluid set of equations of motion [14]. All these equations of motion can be written in the same canonical form (Appendix 1), the form of an Ohm's law [20]

$$\vec{\Sigma}_\sigma + \vec{v}_\sigma \times \vec{\Omega}_\sigma = \vec{R}_\sigma \qquad (2)$$

where the canonical force-field $\vec{\Sigma}_\sigma \equiv -\nabla h_\sigma - \partial \vec{P}_\sigma/\partial t$ represents the conservative and inductive forces and the second term represents the forces that do no work (Coriolis and Lorentz forces). At this point, we are only considering the fluid regime, so the characteristic velocity $\vec{v}_\sigma$ is the bulk fluid velocity of the species $\vec{u}_\sigma$ (i.e. $\vec{v}_\sigma = \vec{u}_\sigma$). Section IV will show that the single particle and Vlasov/Boltzmann equations can also be written in the form of Eq. 2. Section V will generalize Eq. 2 to relativistic regimes by deriving it from a Lagrangian-Hamiltonian point-of-view. The canonical force-field is defined analogously to the electric field so enthalpy $h_\sigma$ and canonical momentum $\vec{P}_\sigma$ can be interpreted as scalar and vector potentials for the dynamics of the system, respectively. This analogy is taken further by showing that $\vec{\Sigma}_\sigma$ and $\vec{\Omega}_\sigma$ obey field equations.

## III. DERIVING THE FIELD EQUATIONS FROM THE CANONICAL EQUATION OF MOTION

Showing that a general form of Maxwell's equations governs $\vec{\Sigma}_\sigma$ and $\vec{\Omega}_\sigma$ reinforces the analogy between canonical force-fields and electromagnetic fields. One procedure is as follows: taking the circulation of the canonical force-field produces Faraday's law for canonical quantities $\nabla \times \vec{\Sigma}_\sigma = -\partial \vec{\Omega}_\sigma/\partial t$; taking the divergence of the canonical vorticity produces Gauss' law for solenoidal fields $\nabla \cdot \vec{\Omega}_\sigma = 0$; taking the divergence of Eq. 2 produces Gauss' law for the canonical force-field, $\nabla \cdot \vec{\Sigma}_\sigma = \mathbb{q}_\sigma/\epsilon_\sigma$, provided the source $\mathbb{q}_\sigma$ is defined as

$$\frac{\mathbb{q}_\sigma}{\epsilon_\sigma} \equiv \nabla \cdot [\vec{R}_\sigma - \vec{v}_\sigma \times \vec{\Omega}_\sigma] \qquad (3)$$

where the constant $\epsilon_\sigma$ remains to be determined; taking the circulation of the canonical vorticity produces Ampère's law for canonical fields $\nabla \times \vec{\Omega}_\sigma = \mu_\sigma \vec{\mathbb{j}}_\sigma + \mu_\sigma \epsilon_\sigma \partial \vec{\Sigma}_\sigma/\partial t$, where the constant $\mu_\sigma$ remains to be determined, provided a canonical current density is defined as $\vec{\mathbb{j}}_\sigma \equiv \mathbb{q}_\sigma \vec{v}_\sigma$, and the divergence of the canonical momentum is related to enthalpy by

$$\nabla \cdot \vec{P}_\sigma = -\mu_\sigma \epsilon_\sigma \frac{\partial h_\sigma}{\partial t}. \qquad (4)$$

Eq. 4 is a Lorenz gauge choice, which states that enthalpy evolution within the system is equivalent to having canonical momentum intercept the system boundaries. We choose to call $\vec{\Sigma}_\sigma$ a force-field because although it has units of force, it behaves like a field by obeying these generalized Maxwell field equations (as does $\vec{\Omega}_\sigma$). The enthalpy $h_\sigma$ and canonical momentum $\vec{P}_\sigma$ therefore act as scalar and vector potentials for the canonical fields $\vec{\Sigma}_\sigma$ and $\vec{\Omega}_\sigma$. The same procedures as the ones for electromagnetic fields gives inhomogeneous wave equations for these $h_\sigma$ and $\vec{P}_\sigma$ potentials and a Poynting theorem for these $\vec{\Sigma}_\sigma$ and $\vec{\Omega}_\sigma$ fields.

## IV. DERIVING CANONICAL OHM'S LAW FROM SINGLE-PARTICLE AND KINETIC EQUATIONS OF MOTION

To move beyond the fluid regime, this section shows the canonical equation of motion (Eq. 2) and thus the generalized Maxwell's equations (Sec. III) and the helicity transport equation (Eq. 1) are all also valid for single and kinetic distributions of particles. The Vlasov-Boltzmann equation for a velocity distribution $f_\sigma(\vec{v}_\sigma)$ can be re-arranged exactly (Appendix 2) into the form of Eq. 2, $\vec{\Sigma}_\sigma^{kin} + \vec{v}_\sigma \times \vec{\Omega}_\sigma^{kin} = \vec{R}_\sigma^{kin}$, provided the frictional term is defined as

$$\vec{R}_\sigma^{kin} \equiv \left(\vec{v}_\sigma \cdot \vec{P}_\sigma^{par} - h_\sigma^{par}\right) \nabla f_\sigma - \vec{P}_\sigma^{par} \frac{df_\sigma}{dt}. \qquad (5)$$

The labels $par$, $kin$, and $flu$ specify the particle, kinetic and fluid regime, respectively (Table 1). The collision operator, reaction sources and velocity space terms of Vlasov's equation are implicit inside the factor $df/dt = \partial f/\partial t + \vec{v} \cdot \nabla f$ on the right-hand side of Eq. 5. The symbol $\vec{v}_\sigma$ now represents the velocity of the kinetic group of particles, whereas earlier, the symbol represented the bulk fluid velocity $\vec{u}_\sigma$.

Setting $f_\sigma = 1$ retrieves the single particle equation of motion re-arranged into the form of Eq. 2,

$\vec{\Sigma}_\sigma^{par} + \vec{v}_\sigma \times \vec{\Omega}_\sigma^{par} = 0$, where the symbol $\vec{v}_\sigma$ now represents the single particle velocity. One can also begin with the single particle equation of motion and re-arrange it directly into this form (Appendix 3). At first glance, it appears that for a single particle $\vec{R}_\sigma^{par} = 0$, i.e. there are no dissipative forces, but the exact more fundamental form will be presented below. Finally, to complete the circle, it is possible with some work to retrieve the fluid version of Eq. 2 from the kinetic version of Eq. 2: separating the velocity $\vec{v} = \vec{u} + \vec{v}'$ into the bulk fluid component $\vec{u}$ and random component $\vec{v}'$ results in an intermediate form of Eq. 2 with pairs of terms representing the fluid component and the random components. Integrating over velocity space then results in the fluid form of Eq. 2, $\vec{\Sigma}_\sigma^{flu} + \vec{u}_\sigma \times \vec{\Omega}_\sigma^{flu} = \vec{R}_\sigma^{flu}$, provided the dissipative term is

$$\vec{R}_\sigma^{flu} \equiv \vec{R}_{\sigma\alpha} + \vec{R}_{\sigma\sigma} + \vec{R}_{\sigma c} + \vec{R}_{\sigma n}. \quad (6)$$

The four non-conservative forces are the frictional forces $\vec{R}_{\sigma\alpha} \equiv -\sum_\alpha \int m_\sigma \vec{v}_\sigma\, C_{\sigma\alpha}(f_\sigma)\, dv_\sigma$ due to the inter-species collision operator $C_{\sigma\alpha}$, the viscous forces $\vec{R}_{\sigma\sigma} \equiv \nabla \cdot (\Pi_\sigma - \mathcal{P}_\sigma \hat{I})$ resulting from the off-diagonal terms of the pressure tensor $\Pi_\sigma \equiv \int f_\sigma m_\sigma \vec{v}'_\sigma \vec{v}'_\sigma\, d\vec{v}'_\sigma$ away from the scalar pressure $\mathcal{P}_\sigma \equiv \int f_\sigma m_\sigma v_\sigma'^2/2\, dv'_\sigma$, the compressible effects $\vec{R}_{\sigma c} \equiv (\rho_\sigma \vec{u}_\sigma + \rho_{c\sigma} \vec{A}) \nabla \cdot \vec{u}_\sigma$, and the density gradient effects $\vec{R}_{\sigma n} \equiv (-m_\sigma \phi + m_\sigma u_\sigma^2/2 + m_\sigma \vec{u}_\sigma \cdot \vec{A}) \nabla n_\sigma$. The unit tensor is $\hat{I}$. Here, the fluid canonical momentum is $\vec{P}_\sigma^{flu} = \rho_\sigma \vec{u}_\sigma + \rho_{c\sigma} \vec{A}$ and the fluid enthalpy is $h_\sigma^{flu} = \rho_{c\sigma} \phi + \rho_\sigma u_\sigma^2/2 + \mathcal{P}_\sigma$ where the fluid parcel has a mass density $\rho_\sigma$ and charge density $\rho_{c\sigma}$. Our earlier derivation [20] of the fluid form of Eq. 2 began with the fluid not the kinetic equation and used $\vec{P}_\sigma^{flu}/n_\sigma$, $h_\sigma^{flu}/n_\sigma$, and $\vec{R}_{\sigma n} = 0$ with a barotropic assumption. Eq. 6 is more general and reverts to the earlier form when dividing by $n_\sigma$ and eliminating terms with a sourceless continuity equation. Canonical momentum $\vec{P}$, enthalpy $h$, canonical force-field $\vec{\Sigma}$, and canonical vorticity $\vec{\Omega}$ are thus generalizations, not just analogs, of magnetic potential $\vec{A}$, electrostatic potential $\phi$, electric field $\vec{E}$, and magnetic field $\vec{B}$, respectively. A more fundamental approach generalizes the idea further, determines the constants, and suggests possible interpretations.

**V.  LAGRANGIAN-HAMILTONIAN FORMALISM**

We can now derive the canonical equation of motion (Eq. 2) and the associated field equations (Sec. III) from a more fundamental point-of-view for all three regimes (single particle, kinetic, and fluid). This provides more insight into the nature of the canonical quantities, it further generalizes Eq. 2 to relativistic conditions with sources and sinks, and it provides the basis for comparing helicity with energy conservation. Inspired by the Maxwell Lagrangian [23], we define a new species Lagrangian density in four-dimensional notation $x^\mu = \{ct, x, y, z\}$ as

$$\mathcal{L}_\sigma \equiv \mathbb{J}^\nu \mathbb{P}_\nu - \frac{1}{4\mu_\sigma} \mathbb{F}_{\mu\nu} \mathbb{F}^{\mu\nu} \qquad (7)$$

where the canonical form of the four-current is $\mathbb{J}^\mu \equiv \mathbb{q}\, \partial_t x^\mu$, the four-potential is $\mathbb{P}^\mu \equiv \{h/c\,; \vec{P}\}$, and the four-field tensor is $\mathbb{F}_{\mu\nu} \equiv \partial_\mu \mathbb{P}_\nu - \partial_\nu \mathbb{P}_\mu$. The generalized charge density $\mathbb{q}$ is defined in Eq. 3. The symbol $\partial_\mu$ represents the derivative operator in the chosen manifold with $\partial_t$ being the shorthand for the time component. For clarity, the species subscript $\sigma$ is dropped for four-dimensional notation but, for consistency, it is kept for other symbols. The constant $c$ represents the speed of light and the Einstein summation rules apply. This Lagrangian has the same form (Eq. 7) for all regimes provided the enthalpy and canonical momentum are $h_\sigma^{par} = \gamma m_\sigma c^2 + q_\sigma \phi$ and $\vec{P}_\sigma^{par} = \gamma m_\sigma \vec{v}_\sigma + q_\sigma \vec{A}$ for single particles ($\gamma$ is the Lorentz factor), $h_\sigma^{kin} = f_\sigma h_\sigma^{par}$ and $\vec{P}_\sigma^{kin} = f_\sigma \vec{P}_\sigma^{par}$ for kinetic distributions, and $h_\sigma^{flu} = \int h_\sigma^{kin} d\vec{v}_\sigma$ and $\vec{P}_\sigma^{flu} = \int \vec{P}_\sigma^{kin} d\vec{v}_\sigma$ for fluids, respectively (Table 1). The classical forms are retrieved to an inconsequential constant in a flat metric with the weak gravitational field $\phi_g \ll c^2$ and slow motion $v \ll c$ approximations so $\gamma \simeq 1 + v^2/(2c^2) + \phi_g/c^2$. The fluid approximation also assumes that the bulk flow $\vec{u}_\sigma$ can be relativistic while the random component $\vec{v}'_\sigma$ remains classical, so $\gamma \to \gamma(u_\sigma)$. Inserting Eq. 7 into the Euler-Lagrange equation

$$\frac{\partial \mathcal{L}}{\partial y} - \sum_{i=0}^{N} \frac{\partial}{\partial q_i}\left(\frac{\partial \mathcal{L}}{\partial(\partial_{q_i} y)}\right) = 0 \qquad (8)$$

for the canonical four-potential $y \to \mathbb{P}^\nu$ as a function of the four-coordinates $q_i \to x^\mu$ with $N = 1$ gives

Maxwell's equation for the canonical quantities

$$D_\mu \mathbb{F}^{\mu\nu} = -\mu_\sigma \mathbb{J}^\nu \qquad (9)$$

where $\partial_\mu$ has been replaced by the covariant derivative $D_\mu$ via the equivalence principle [24]. The canonical four-potential wave equation $\partial_\mu \partial^\mu \mathbb{P}^\nu = -\mu_\sigma \mathbb{J}^\nu$ is then obtained upon substitution of the four-field tensor in Eq. 9 by its definition and choosing the Lorenz gauge $\partial_\mu \mathbb{P}^\mu = 0$, which is equivalent to Eq. 4 in flat space when $c^2 = 1/\mu_\sigma \epsilon_\sigma$. Inserting Eq. 7 into the Euler-Lagrange equation (Eq. 8) for the four-coordinates $y \to x^\mu$ as a function of time $q_i \to t$ with $N = 1$ gives an Ohm's law for canonical quantities, i.e. a transport equation of the canonical four-potential

$$\mathbb{q}\left[\partial_t \mathbb{P}_\mu - \partial_t x^\nu\, \partial_\mu \mathbb{P}_\nu\right] = -\partial_\mu\left(\frac{1}{4\mu_\sigma} \mathbb{F}_{\alpha\beta} \mathbb{F}^{\alpha\beta}\right). \qquad (10)$$

In the Minkowski metric, $\mathbb{J}^\nu \mathbb{P}_\nu = \vec{\mathbb{J}}_\sigma \cdot \vec{P}_\sigma - \mathbb{q}\, h_\sigma$ and $-\mathbb{F}_{\mu\nu}\mathbb{F}^{\mu\nu}/4\mu_\sigma = \epsilon_\sigma \Sigma_\sigma^2/2 - \Omega_\sigma^2/2\mu_\sigma$, so Eq. 7 can also be written as differences between generalized kinetic energies and potential energies, $\mathcal{L}_\sigma = \vec{\mathbb{J}}_\sigma \cdot \vec{P}_\sigma - \mathbb{q}\, h_\sigma + \epsilon_\sigma \Sigma_\sigma^2/2 - \Omega_\sigma^2/2\mu_\sigma$. In this metric, Eq. 10 for the index $\mu = 0$ retrieves the energy equation

$$\mathbb{q}\left[\vec{v}_\sigma \cdot \frac{\partial \vec{P}_\sigma}{\partial t}\right] = -\frac{\partial}{\partial t}\left(\frac{1}{2}\epsilon_\sigma \Sigma_\sigma^2 - \frac{\Omega_\sigma^2}{2\mu_\sigma}\right) \qquad (11)$$

and for $\mu = j \in \{1, 2, 3\}$, Eq. 10 retrieves the full canonical equation of motion

$$\mathbb{q}\left[\vec{\Sigma}_\sigma + \vec{v}_\sigma \times \vec{\Omega}_\sigma\right] = \vec{R}_\sigma \qquad (12)$$

provided

$$\vec{R}_\sigma \equiv -\nabla\left(\frac{1}{2}\epsilon_\sigma \Sigma_\sigma^2 - \frac{\Omega_\sigma^2}{2\mu_\sigma}\right). \qquad (13)$$

Eq. 12 generalizes Eq. 2 to situations when $\mathbb{q} \neq 1$, i.e. non-conventional matter as a force-field source given by Eq. 3 and expressed as a canonical four-current $\mathbb{J}^\mu$ (e.g. refs. 25 and 26 on non-Abelian fluids). When $\mathbb{q} = 0$, Eq. 9 is source-free and Eq. 7 only retains the coupling between canonical field components.

Eq. 13 shows that dissipative forces represent an incomplete conversion of canonical vorticity "potential" energy $\Omega_\sigma^2/2\mu_\sigma$ into canonical force-field "kinetic" energy $\epsilon_\sigma \Sigma_\sigma^2/2$. For single particles, the canonical vorticity energy is just the magnetic energy, $\Omega_\sigma^{par\,2}/2\mu_\sigma = q_\sigma^2 B^2/2\mu_\sigma$ because velocity and position are independent in phase space, and the canonical force-field energy is just the electric field energy, $\epsilon_\sigma \Sigma_\sigma^{2\,par}/2 \simeq \epsilon_\sigma q_\sigma^2 E^2/2$ in the weak gravitational field and slow motion approximation. These give $\mu_\sigma = \mu_0 q_\sigma^2$ and $\epsilon_\sigma = \epsilon_0/q_\sigma^2$ where $\mu_0, \epsilon_0$ are the vacuum permeability and permittivity, respectively. These definitions also satisfy the Lorenz gauge choice above. These approximations are equivalent to considering that there are no sources or sinks in the electromagnetic field, i.e. that the presence of the particle does not affect the field, so $B^2/2\mu_0 = \epsilon_0 E^2/2$ and $\Omega_\sigma^{par\,2}/2\mu_\sigma = \epsilon_\sigma \Sigma_\sigma^{2\,par}/2$, therefore Eq. 13 gives $\vec{R}^{par} \simeq 0$ as before. In the case of strong gravitational fields or relativistic motion, $\vec{R}^{par} \neq 0$, corresponding to a finite non-electric term $\vec{\Sigma}_\sigma^{\prime\,par} = -\nabla(\gamma m_\sigma c^2) - \partial(\gamma m_\sigma \vec{v}_\sigma)/\partial t$ in the expansion $\vec{\Sigma}_\sigma^{par} = q_\sigma \vec{E} + \vec{\Sigma}_\sigma^{\prime\,par}$. In this case, Eqs. 3, 7, and 13 show that particle motion acts as source terms for the canonical fields (e.g. synchrotron, bremmstrahlung, gravitational emission). Finally, substituting the Newtonian kinetic or fluid forms for $\vec{\Sigma}_\sigma, \vec{\Omega}_\sigma$ into Eq. 13, and using Hamilton's equation to give $\nabla \mathcal{L}_\sigma = d\vec{P}_\sigma/dt$ or re-defining collisional friction as random accelerations $\vec{R}_{\sigma\alpha} = \int f_\sigma m_\sigma d\vec{v}_\sigma'/dt\, d\vec{v}_\sigma'$ retrieves Eq. 5 or 6, respectively. This consistency reinforces the choice of Lagrangian (Eq. 7). Collective behavior due to $\nabla_v \cdot (f_\sigma \vec{P}^{par})$ is then represented as finite non-conservative forces in the kinetic version of Eqs. 12 and 13. By explicitly expanding the canonical force-field into the electric field and motion components, $\vec{\Sigma}_\sigma^{kin} = f_\sigma q_\sigma \vec{E} + \vec{\Sigma}_\sigma^{\prime\,kin}$, and the canonical vorticity into the magnetic field and vortex components, $\vec{\Omega}_\sigma^{kin} = f_\sigma q_\sigma \vec{B} + \vec{\Omega}_\sigma^{\prime\,kin}$, where $\vec{\Sigma}_\sigma^{\prime kin} = f_\sigma \vec{\Sigma}_\sigma^{\prime\,par} - h_\sigma^{par} \nabla f_\sigma - \vec{P}^{par} \partial f_\sigma/\partial t$ and $\vec{\Omega}_\sigma^{\prime\,kin} = \nabla f_\sigma \times \vec{P}_\sigma^{par}$, Eq. 7 can be written as $\mathcal{L}_\sigma = \mathcal{L}_s + \mathcal{L}_M + \mathcal{L}_\sigma'$ provided (in flat space)

$$\mathcal{L}_\sigma' = f_\sigma q_\sigma \left( \epsilon_\sigma \vec{E} \cdot \vec{\Sigma}_\sigma^{\prime\,kin} - \frac{\vec{B} \cdot \vec{\Omega}_\sigma^{\prime\,kin}}{\mu_\sigma} \right) + \frac{1}{2} \epsilon_\sigma \Sigma^{\prime\,kin\,2} - \frac{\Omega_\sigma^{\prime\,kin\,2}}{2\mu_\sigma}. \qquad (14)$$

Here, $\mathcal{L}_s = \vec{j} \cdot \vec{P}_\sigma - \mathbb{q} h_\sigma$ and $\mathcal{L}_M = \epsilon_0 E^2/2 - B^2/2\mu_0$ are the usual source and Maxwell Lagrangians in Minkowski space, respectively, if the charge $\mathbb{q}$ defined by Eq. 3 is unity and the associated current $\vec{j} = \mathbb{q}\vec{v} = \vec{v}$. In the kinetic case, the $\epsilon_\sigma, \mu_\sigma$ coefficients become $\epsilon_\sigma = \epsilon_0/(f_\sigma^2 q_\sigma^2)$ and $\mu_\sigma = f_\sigma^2 q_\sigma^2 \mu_0$ respectively. Eq. 14 represents the collective coupling between kinetic distributions and electromagnetic fields, between distributions themselves, and relativistic contributions. For non-relativistic single particles,

Eq. 14 vanishes, retrieving the textbook Lagrangian forms. Because Eq. 14 is Maxwell-like, these interactions can be interpreted with Eq. 9 on non-electromagnetic components (e.g. three-wave interactions, ray or Gaussian optics, etc.) and Eq. 1 for constrained relaxation [27].

## VI. HELICITY EVOLUTION COMPARED TO ENERGY EVOLUTION

Constrained relaxation rests on the hypothesis that the constraint (helicity) is fixed when the system evolves to a lower energy state. This section briefly compares the helicity evolution to the energy evolution. The Legendre transformation of Eq. 7 for two canonical variables $\mathbb{P}_\mu$ and $x_\mu$ gives the Hamiltonian density of the species

$$\mathcal{H}_\sigma = \frac{1}{2}\epsilon_\sigma \Sigma_\sigma^2 + \frac{\Omega_\sigma^2}{2\mu_\sigma} + \mathbb{q}\, h_\sigma \qquad (15)$$

(in flat space) where an extra term $-\epsilon_\sigma \nabla \cdot (h_\sigma \vec{\Sigma}_\sigma)$ has been set to zero. This offset term is ignored, assuming it integrates away when calculating the total Hamiltonian $H = \int \mathcal{H}\, dV$ over the whole volume (otherwise reference fields would be needed). Eq. 15 gives the sum of all the forms of energy in the system, i.e. the field energy and the source enthalpy, and when inserted into Hamilton's equation $d\vec{P}_\sigma/dt = -\nabla \mathcal{H}_\sigma$ provides an alternative derivation of Eq. 12. Using the Poynting theorem for canonical quantities $\vec{\Sigma}_\sigma, \vec{\Omega}_\sigma$ in the Hamilton equation $\partial \mathcal{H}/\partial t = -\partial \mathcal{L}/\partial t$ and integrating over the volume gives the energy evolution equation

$$\frac{dH_\sigma}{dt} = \int \mathbb{q}\frac{\partial h_\sigma}{\partial t} dV - \int \mathbb{q}\, \vec{\Sigma}_\sigma \cdot \vec{v}_\sigma\, dV - \int \frac{\vec{\Sigma}_\sigma \times \vec{\Omega}_\sigma}{\mu_\sigma} \cdot d\vec{S} + \int \mathcal{H}_\sigma \vec{u}_\sigma \cdot d\vec{S}. \qquad (16)$$

The terms represent changes of enthalpy within the volume, the energy changes due to motion, the Poynting flux of canonical fields out of the volume, and changes due to motion of the boundary, respectively. Comparing Eq. 1 with Eq. 16 then provides the conditions for helicity evolution with respect to energy evolution.

For example, consider the simplest case of a fixed, isolated system that is subject to dissipative forces. Substituting Eqs. 4 and 11 into Eq. 16, then over a time increment, the ratio of Eq. 1 to Eq. 16 simplifies to

$$\frac{\Delta K_\sigma/K_{\sigma 0}}{\Delta H_\sigma/H_{\sigma 0}} = 2\frac{\int \vec{R}_\sigma \cdot \vec{\Omega}_\sigma \, dV}{\int \vec{R}_\sigma \cdot \vec{v}_\sigma \, dV}\frac{H_{\sigma 0}}{K_{\sigma 0}} \simeq 2\frac{\Omega_\sigma}{v_\sigma}\frac{H_{\sigma 0}}{K_{\sigma 0}} \qquad (17)$$

where canonical helicity and energy have been normalized to the total initial helicity $K_{\sigma 0}$ given by Eq. 1 and the total initial energy $H_{\sigma 0}$ given by Eq. 14, respectively. Eq. 17 shows that the ratio of canonical vorticity to species velocity, $\Omega_\sigma/v_\sigma$, determines whether helicity changes are larger or smaller than energy changes. Taking measured values from the MST experiment [28], assuming the total energy is only magnetic $H_{e0} \sim W_{mag} \sim 50$ kJ, the canonical vorticity is only magnetic $\Omega_e \sim \rho_{ce}B \sim 0.2$, at a plasma density of $n_e \sim 10^{19}$ m$^{-3}$, and a toroidal magnetic field of 0.12 T, a velocity $v_e \sim v_{The} \sim 10^7$ m/s for $T_e \sim 300$ eV, , and a magnetic helicity of 23 mWb$^2$, into Eq. 17 shows that magnetic energy is dissipated 33 times more than magnetic helicity in an isolated, purely dissipative, magnetically-dominated system. For more complex scenarios involving multi-scale effects across several regimes (e.g. magnetic reconnection in sawteeth crash), the evolution of canonical helicity and total energy will be governed by Eq. 1 and Eq. 17. Ref. 20 demonstrated that the total canonical helicity is conserved, $\mathbb{K} = K_i + K_e$, and there can be conversion of one type of helicity into another (e.g. magnetic into cross or kinetic helicity, or vice-versa). For fluid regimes, the factor $2\,\Omega_\sigma/v_\sigma$ approximates to a density gradient scale $\rho_\sigma/L_s$ where the scale length is $1/L_s \equiv 2(1/L_{circ} + 1/r_L)$, defined to be a hybrid length of a characteristic Larmor radius $r_L(u_\sigma)$ and the vorticity scale length $L_{circ}$. In neutral fluids or un-magnetized plasmas, canonical helicity changes are determined by the flow circulation scale length, while changes in magnetized plasmas are determined by the hybrid scale length. For kinetic regimes, $\rho_\sigma$ is replaced by $f_\sigma m_\sigma$ and the gradient of the distribution function plays the critical role. These results also apply to collisionless regimes ($\vec{R}_{\sigma\alpha} = 0$), inviscid regimes ($\vec{R}_{\sigma\sigma} = 0$), and incompressible regimes ($\vec{R}_{\sigma c} = 0$) provided a density gradient exists ($\vec{R}_{\sigma n} \neq 0$). For single particles, Eq. 17 approaches zero (helicity is conserved) because generally $m_\sigma H_{\sigma 0}/(L_s K_{\sigma 0}) \ll 1$, unless scale lengths are small, of $\mathcal{O}(m_\sigma H_{\sigma 0}/K_{\sigma 0})$. This is another expression of adiabatic invariance. For example, for ion with $m_i \sim 10^{-27}$ kg and $q_i \sim 10^{-19}$ C, with energy dominated by magnetic energy $H_i \sim B^2/2\mu_0$, a canonical helicity dominated by magnetic helicity $K_i = q_i^2 K_{mag} \sim q_i^2 (B \pi r_{Li})^2$ with Larmor radius of $r_{Li} \sim 1$ cm typical for magnetic confinement experiments, the coefficient $m_i H_{i0}/(r_{Li} K_{i0}) \sim 10^{-12}$, confirming that magnetic helicity is a good constant of the motion for this single ion.

## VII. SUMMARY

This paper developed a unifying field-theory framework for the dynamics of single-particle, kinetic and fluid models. The framework demonstrates that a species' canonical helicity evolution is not limited to Newtonian fluid regimes. The framework also shows that a species' canonical helicity is well conserved compared to the species' energy in shallow density gradients but not in steep density gradients (in the simplest case of an isolated, dissipative system). These results suggest that in the edge of multi-species, collisionless, kinetic plasmas, magnetic helicity can couple to ion canonical helicity, spontaneously generating flowing structures when density gradients are of the order of the ion skin depth [20]. These regimes apply to magnetic fusion plasmas. A cosmic dynamo operates in reverse, with gravitational potential energy acting as an enthalpy source of ion canonical helicity that couples to magnetic helicity, spontaneously generating magnetic structures, when density gradients approach hybrid scale lengths. These ideas will be detailed in future work. This field theory approach to helicity and energy evolution suggests that techniques borrowed from electromagnetism would be suitable for analyzing self-organization and magnetic reconnection, similar to how existing analysis of helicity injection into toroidal magnetic configurations uses electrical circuit analysis techniques [12]. On the largest scale, adding Einstein's Lagrangian to Eq. 7 and calculating the circulation of Eq. 10 to give curved manifold version of Eqs. 1 and 16 would extend the concept of self-organization by helicity-constrained relaxation to general relativity. This has applications, amongst others, to astrophysical jet formation from accretion disks around super-massive black holes in active galactic nuclei. The unifying framework can also be useful because there is a one-to-one relationship between $\vec{\Omega}_\sigma$ and the plasma species at all regimes, and MHD intuition can be used for all plasma regimes, provided one replaces the magnetic field $\vec{B}$ with the canonical vorticity $\vec{\Omega}_\sigma$, electrostatic potential and pressure with a general enthalpy $h_\sigma$, and the electric field $\vec{E}$ with the canonical force-field $\vec{\Sigma}_\sigma$.

**Acknowledgements**

This work was supported by the U.S. Dept. of Energy Grant DE-SC0010340. The author would like to acknowledge one of the referees for pointing out Refs. 18 and 19. These earlier works partly explored some of the connections presented here.

configurations (Eq. 1 with $m_\sigma \to 0$). In particular, electrical circuit methods were used to analyze such configurations. Concepts of alternating or direct currents, voltages, impedances, resonances, etc. were useful and practical because a magnetostatic plasma is an electrical load for the driving circuits. The Maxwell form of the unifying Lagrangian (Eq. 7) confirms that electrical engineering methods are useful for magnetostatic plasmas; and suggests that such methods should also be useful for flowing magnetized (or non-magnetized) plasmas and flowing neutral fluids with finite vorticity. The driving circuits can be any combination of gravitational, pressure, kinetic or electrical supplies since these power supplies are simply enthalpy sources for a canonical Maxwell cicuit.

[28] H. Ji, S. C. Prager and J. S. Sarff, Phys. Rev. Lett., 74, 2945-2948, (1995).

**APPENDIX 1: DERIVATION OF CANONICAL EQUATION OF MOTION FROM FLUID EQUATION OF MOTION**

Begin with the fluid equation of motion for a species $\sigma$ with mass density $\rho_\sigma$ and charge density $\rho_{c\sigma}$

$$\rho_\sigma \frac{\partial \vec{u}_\sigma}{\partial t} + \vec{u}_\sigma \cdot \nabla \vec{u}_\sigma = \rho_{c\sigma}(\vec{E} + \vec{u}_\sigma \times \vec{B}) - \nabla \mathcal{P} - \vec{R}_{\sigma\sigma} - \vec{R}_{\sigma\alpha} \quad (18)$$

where the $\nabla = \partial/\partial \vec{x}$ represents the differential operator in space, $\mathcal{P}$ represents the pressure, $\vec{R}_{\sigma\sigma}$ represents viscosity, and $\vec{R}_{\sigma\alpha}$ represents interspecies collisions. Expanding the second term using $\vec{E} = -\nabla \phi - \partial \vec{A}/\partial t$ and using appropriate vector identities on the second terms of the left-hand side and the right-hand side gives

$$\frac{\partial(\rho_\sigma \vec{u}_\sigma + \rho_{c\sigma}\vec{A})}{\partial t} - \vec{u}_\sigma \times \left[\nabla \times (\rho_\sigma \vec{u}_\sigma + \rho_{c\sigma}\vec{A})\right]$$

$$= -\nabla\left[\rho_{c\sigma}\phi + \frac{\rho_\sigma u_\sigma^2}{2} + \mathcal{P}\right] - \vec{R}_{\sigma\sigma} - \vec{R}_{\sigma\alpha}$$

$$+ \left\{\vec{u}_\sigma \frac{\partial \rho_\sigma}{\partial t} + \frac{u_\sigma^2}{2}\nabla \rho_\sigma - \vec{u}_\sigma \times (\nabla \rho_\sigma \times \vec{u}_\sigma) + \phi \nabla \rho_{c\sigma} + \vec{A}\frac{\partial \rho_c}{\partial t} - \vec{u}_\sigma \right.$$

$$\left. \times (\nabla \rho_{c\sigma} \times \vec{A})\right\}. \quad (19)$$

Using vector identities on the sixth and last terms on the right-hand side simplifies the group inside the curly brackets to

$$\{\ldots\} = \vec{u}_\sigma\left[\frac{\partial \rho_\sigma}{\partial t} + \vec{u}_\sigma \cdot \nabla \rho_\sigma\right] - \frac{u_\sigma^2}{2}\nabla \rho_\sigma + \phi \nabla \rho_{c\sigma} + \vec{A}\left[\frac{\partial \rho_{c\sigma}}{\partial t} + \vec{u}_\sigma \cdot \nabla \rho_{c\sigma}\right] - \vec{u}_\sigma \cdot \vec{A}\nabla \rho_{c\sigma} \quad (20)$$

The continuity equation reduces these square brackets to $-\rho_\sigma \nabla \cdot \vec{u}_\sigma$ and $-\rho_{c\sigma}\nabla \cdot \vec{u}_\sigma$, respectively, (source terms can be included here if necessary) which simplifies Eq. 18 to

$$\frac{\partial \vec{P}_\sigma^{flu}}{\partial t} - \vec{u}_\sigma \times [\nabla \times \vec{P}_\sigma^{flu}]$$

$$= -\nabla h_\sigma^{flu} - \vec{R}_{\sigma\sigma} - \vec{R}_{\sigma\alpha} - \vec{P}_\sigma^{flu}\nabla \cdot \vec{u}_\sigma \quad (21)$$

$$- \left(-q_\sigma\phi + \frac{1}{2}m_\sigma u_\sigma^2 + q_\sigma \vec{u}_\sigma \cdot \vec{A}\right)\nabla n.$$

using the definitions of Table 1. The compressibility term on the right-hand side can be labelled $\vec{R}_{\sigma c} = \vec{P}_\sigma^{flu}\nabla \cdot \vec{u}_\sigma$ and the density gradient term can be labelled $\vec{R}_{\sigma n} = \left(-q_\sigma\phi + \frac{1}{2}m_\sigma u_\sigma^2 + q_\sigma \vec{u}_\sigma \cdot \vec{A}\right)\nabla n$. Using the definition of the canonical force-field $\vec{\Sigma} \equiv -\nabla h - \partial \vec{P}/\partial t$ and defining the dissipative term $\vec{R}_\sigma = \vec{R}_{\sigma\alpha} + \vec{R}_{\sigma\sigma} + \vec{R}_{\sigma c} + \vec{R}_{\sigma n}$ (i.e. Eq. 6) allows us to write the fluid Eq. 19 in the canonical form

$$\vec{\Sigma}_\sigma^{flu} + \vec{u}_\sigma \times \vec{\Omega}_\sigma^{flu} = \vec{R}_\sigma^{flu}. \quad (22)$$

This derivation is included here for convenience but has been presented elsewhere for neutral fluids [3, 2], for magnetohydrodynamics [15], for multi-fluids [14], and for gauge-invariant relative helicity [20]. Generalizing this to relativity and regimes beyond the fluid regimes requires the more formal Lagrangian approach (Sec. V).

**APPENDIX 2: DERIVATION OF CANONICAL EQUATION OF MOTION FROM KINETIC EQUATION OF MOTION**

Begin with the Vlasov-Boltzmann equation

$$\frac{\partial f_\sigma}{\partial t} + \vec{v}_\sigma \cdot \nabla f_\sigma + \nabla_v \cdot (f_\sigma \vec{a}) = C(f_\sigma) \quad (23)$$

where $\nabla = \partial/\partial \vec{x}$ represents the differential operator in space; $\nabla_v = \partial/\partial \vec{v}$ the one in velocity space; $\vec{a} = \partial \vec{v}/\partial t$ represents the acceleration term, which can be zero for neutral fluids or the Lorentz $m\vec{a} = q\vec{E} + q\vec{v} \times \vec{B}$ for plasmas; and $C(f_\sigma)$ a general inhomogeneous term. Multiplying by $m_\sigma \vec{v}_\sigma$ and expanding out the terms with vector identities gives

$$\frac{\partial (fm\vec{v})}{\partial t} - fm\vec{a} + \nabla \cdot (fm\vec{v}\vec{v}) + \nabla_v \cdot (fm\vec{a}\vec{v}) - fm\vec{a} \cdot \nabla_v \vec{v} = m\vec{v}C. \quad (24)$$

The subscript $\sigma$ is dropped for clarity. Expanding the second term using $\vec{E} = -\nabla\phi - \partial \vec{A}/\partial t$ and using appropriate vector identities gives

$$fm\vec{a} = -\nabla(fq\phi) + q\phi\nabla f - \frac{\partial(fq\vec{A})}{\partial t} + q\vec{A}\frac{\partial f}{\partial t} + \vec{v} \times [\nabla \times (fq\vec{A})] - q\vec{v} \cdot \vec{A}\,\nabla f + (\vec{v} \cdot \nabla f)q\vec{A}. \quad (25)$$

Expanding the third term of Eq. 24 with vector identities and remembering that $\vec{x}$ and $\vec{v}$ are independent in phase space gives

$$\nabla \cdot (fm\vec{v}\vec{v}) = \nabla(fmv^2) - \vec{v} \times [\nabla \times (fm\vec{v})] - \nabla\left(\frac{fmv^2}{2}\right) + \frac{mv^2}{2}\nabla f. \qquad (26)$$

Substituting Eqs. 25 and 26 into Eq. 24 gives

$$\frac{\partial(fm\vec{v} + fq\vec{A})}{\partial t} - \vec{v} \times [\nabla \times (fm\vec{v} + fq\vec{A})]$$

$$= -\nabla\left(fq\phi + \frac{fmv^2}{2}\right) - \left(-q\phi + q\vec{v} \cdot \vec{A} + \frac{mv^2}{2}\right)\nabla f \qquad (27)$$

$$+ q\vec{A}\left(\frac{\partial f}{\partial t} + \vec{v} \cdot \nabla f\right) + m\vec{v}C - \nabla_v \cdot (fm\vec{a}\vec{v}) + fm\vec{a}.$$

The last three terms on the right-hand side reduces to $m\vec{v}C - \nabla_v \cdot (fm\vec{a}\vec{v}) + fm\vec{a} = m\vec{v}[C - \nabla_v \cdot (f\vec{a})]$ so the Vlasov-Boltzmann equation becomes

$$\frac{\partial \vec{P}^{kin}}{\partial t} - \vec{v} \times [\nabla \times \vec{P}^{kin}] = -\nabla h^{kin} - \mathcal{L}^{par}\nabla f + \vec{P}^{par}\left(\frac{\partial f}{\partial t} + \vec{v} \cdot \nabla f\right). \qquad (28)$$

where the canonical momentum and enthalpies are defined as in Table 1 and the classical particle Lagrangian is the usual $\mathcal{L}^{par} = -q\phi + q\vec{v} \cdot \vec{A} + \frac{mv^2}{2} = \vec{v} \cdot \vec{P}^{par} - h^{par}$. Using the definition of the canonical force-field $\vec{\Sigma} \equiv -\nabla h - \partial \vec{P}/\partial t$ and defining the dissipative term as in Eq. 5 allows us to write the Vlasov-Boltzmann Eq. 22 in the canonical form (with the species subscripts)

$$\vec{\Sigma}_\sigma^{kin} + \vec{v}_\sigma \times \vec{\Omega}_\sigma^{kin} = \vec{R}_\sigma^{kin}. \qquad (29)$$

This derivation shows that the Vlasov-Boltzmann can be written in canonical form but is valid only for Newtonian regimes. Generalizing to relativistic conditions or accounting for the presence of gravity requires the more fundamental approach of Sec. V.

**APPENDIX 3: DERIVATION OF CANONICAL EQUATION OF MOTION FROM SINGLE PARTICLE EQUATION OF MOTION**

Begin with the single particle equation of motion

$$m_\sigma \frac{\partial \vec{v}_\sigma}{\partial t} + m_\sigma \vec{v}_\sigma \cdot \nabla \vec{v}_\sigma = q_\sigma(\vec{E} + \vec{v}_\sigma \times \vec{B}) \qquad (30)$$

where $\nabla = \partial/\partial \vec{x}$ represents the differential operator. Expanding the second term using $\vec{E} = -\nabla\phi - \partial\vec{A}/\partial t$ and using appropriate vector identitites gives

$$\frac{\partial(m_\sigma \vec{v}_\sigma + q_\sigma \vec{A})}{\partial t} - \vec{v}_\sigma \times [\nabla \times (m_\sigma \vec{v}_\sigma + q_\sigma \vec{A})] = -\nabla\left(q_\sigma\phi + \frac{1}{2}m_\sigma v_\sigma^2\right) \qquad (31)$$

Using the definition of the canonical force-field $\vec{\Sigma} \equiv -\nabla h - \partial \vec{P}/\partial t$ and Table 1 allows us to rewrite the equation of motion in canonical form

$$\vec{\Sigma}_\sigma^{par} + \vec{v}_\sigma \times \vec{\Omega}_\sigma^{par} = 0. \tag{32}$$

This derivation results in the same Eq. 17.2 of Ref. 23 which started from the more fundamental Lagrangian formalism. The formal Lagrangian procedure is used for all three regimes (single particle, kinetic, fluid) in Sec. V.